\documentclass[aps,preprintnumbers,showpacs,showkeys,nofootinbib]{revtex4}
\usepackage{epsfig}
\usepackage{amssymb,amsmath,amsfonts,amsthm,graphicx,psfrag}
\graphicspath{{./Figures/}}                 
\usepackage{array}
\usepackage{picture}
\usepackage{subfig}
\usepackage{graphicx}
\usepackage{textpos}
\usepackage{xcolor}
\usepackage{ulem}

\newcommand{\beq}{\begin{equation}}
\newcommand{\eeq}{\end{equation}}
\newcommand{\bea}{\begin{eqnarray}}
\newcommand{\eea}{\end{eqnarray}}

\newcommand{\ds}{\displaystyle}

\definecolor{forest}{rgb}{0.0, 0.4, 0.0}

\begin{document}

\title{Numerical Study of the Roberge-Weiss Transition.}

\author{V.~G.~Bornyakov}
\affiliation{Institute for High Energy Physics NRC "Kurchatov Institute", 142281 Protvino, Russia \\
National Research Center "Kurchatov Institute" Moscow, Russia}

\author{N.~V.~Gerasimeniuk}
\affiliation{Pacific Quantum Center, Far Eastern Federal University, 690922 Vladivostok, Russia}

\author{V.~A.~Goy}
\affiliation{Pacific Quantum Center, Far Eastern Federal University, 690922 Vladivostok, Russia}

\author{A.~A.~Korneev}
\affiliation{Pacific Quantum Center, Far Eastern Federal University, 690922 Vladivostok, Russia}

\author{A.~V.~Molochkov}
\affiliation{Pacific Quantum Center, Far Eastern Federal University, 690922 Vladivostok, Russia}

\author{A.~Nakamura}
\affiliation{RCNP, Osaka University, Osaka 567-0047, Japan}
\affiliation{Pacific Quantum Center, Far Eastern Federal University, 690922 Vladivostok, Russia}

\author{R.~N.~Rogalyov}
\affiliation{Institute for High Energy Physics NRC ``Kurchatov Institute'', 142281 Protvino, Russia \\
}


\begin{abstract}
We study the Roberge-Weiss phase transition numerically. The phase transition is associated with the discontinuities in the quark-number density at specific values of imaginary quark chemical potential. 
We parameterize the quark number density $\rho_q$ by the polynomial fit function to compute the canonical partition functions.
We demonstrate that this approach provides a good framework for analyzing lattice QCD data at finite density and a high temperature. 
We show numerically that at high temperature, the Lee-Yang zeros lie on the negative real semi-axis provided that the high-quark-number contributions to the grand canonical partition function are taken into account.
These Lee-Yang zeros have nonzero linear density, which signals the Roberge-Weiss phase transition. We demonstrate that this density agrees with the quark number density discontinuity at the transition line.
\end{abstract}


\maketitle

\section{Introduction}
\label{sec:introduction}

Properties of strong-interacting matter at 
nonzero temperature $T$ and quark chemical potential $\mu_q$ have received considerable attention. QCD phase diagram in the $T-\mu_q$ plane is expected to have a rich structure, which can be studied experimentally in 
heavy-ion collisions
and astronomical observations of the neutron stars.
For many years, it had been thought that the lattice-QCD 
simulations at finite 
$\mu_q$ are impossible because of the sign problem.  And yet, thanks to recent developments, now one can access the regions at finite $T$ and $\mu_q$ up to 
$\mu_q/T \sim 2$, see, e.g. \cite{Borsanyi:2022qlh,Bollweg:2022rps}.

There is no the sign problem when the chemical potential is pure imaginary, $\mu_q=i\mu_{qI}$. Therefore, one can evaluate various quantities using standard Monte Carlo simulations. It opens the way to compute the canonical partition functions $Z_C(n, T, V)$ and thus, the grand canonical partition function, $Z_{GC}$,  becomes available at both real and imaginary chemical potential due to the following fugacity expansion 
\begin{equation}
Z_{GC}(\theta, T, V) = \sum_{n=-\infty}^{\infty} Z_C(n,T,V) \xi_B^n, \qquad \xi_B=e^{\mu_B/T}, \qquad  \mu_B=N_c \mu_q,  \qquad \theta={\mu_q\over T}=\theta_R + \imath \theta_I \;.
\label{Eq:FugacityExpansion}
\end{equation}
\noindent We call this the canonical approach.

One needs the quark number density to compute the canonical partition functions, which can be provided either by lattice results~\cite{Bornyakov:2016wld, Bonati:2010gi, Bonati:2014kpa, Philipsen:2019ouy} or by phenomenological models~\cite{Vovchenko:2017gkg, Almasi:2018lok}.  Computation of $Z_C(n, T, V)$ and Lee-Yang zeros (LYZ)~\cite{Lee:1952ig} for these 
and other models can bring more understanding of the QCD phase structure at finite temperature and baryon density.

We presented some results in this direction in Ref.~\cite{Wakayama:2018wkc}, here we report substantial progress. Namely, we compute  $Z_C(n, T, V)$ at a high $T$ above the Roberge-Weiss (RW) transition temperature $T_{RW}$ \cite{RobergeWeiss} where a polynomial function of the (imaginary) quark chemical potential fits lattice results for the (imaginary) quark number density very well. We suggest a new approach to the computation of $Z_C(n, T, V)$, which solves the problem we reported in Ref.~\cite{Wakayama:2018wkc}. We show that this new approximate solution for $Z_C(n, T, V)$ works exceptionally well in the infinite volume limit. 
As was shown in \cite{Bornyakov:2020jjl}, the canonical partition functions  are of phenomenological significance because they are related to the probabilities ${\cal P}_n$ that the net baryon number at given values of $\mu_B$ and $T$ equals $n$:
\beq \label{eq:probabilities}
{\cal P}_n = {Z_C(n,T,V) \xi_B^n \over Z_{GC}(\theta,V,T)}\; .
\eeq
The values of $Z_C(n,T,V)$ by themselves  represent relative probabilities to find net baryon number $n$ at $\mu_B=0$.

Here we employ $Z_C(n, T, V)$ to compute the LYZ. We find that 
most of the LYZ lies on the real negative semiaxis in the complex fugacity $\xi_B$ plane.
We then present numerical evidence that the nonzero linear density of LYZ on the real negative semiaxis corresponds to the RW transition in the infinite-volume limit. 

Let us note that by studying QCD at nonzero imaginary chemical potential, one can gain an understanding not only of the RW transition quite intensively investigated in lattice QCD (see e.g.~\cite{Bonati:2010gi, Bonati:2014kpa, Philipsen:2019ouy, Nagata:2014fra, Cardinali:2021fpu}) but also of the critical properties of QCD related to deconfinement, chiral symmetry restoration, and critical endpoint \cite{deForcrand:2002hgr, DElia:2002tig, DElia:2007bkz, Almasi:2018lok, Wakayama:2020dzz}.

The paper is organized as follows. In Section~\ref{sec:headings} we describe the numerical procedure to compute $Z_C(n, T, V)$ as well as their asymptotic estimate, both based on the use of the polynomial fit to our lattice data for the quark number density. In Section~\ref{sec:others} we compute the LYZ and discuss their distribution pattern in the fugacity plane. The Conclusions section summarizes the obtained results.

\section{Computation of the canonical partition functions}
\label{sec:headings}

We simulate $N_s^3 \times 4$ lattices with 
$N_s=16, 20, 40$ at temperature $T/T_c=1.35$ with
$m_{\pi}/m_{\rho}=0.8$. As in Refs.~\cite{Bornyakov:2016wld,Wakayama:2018wkc}
we use the lattice QCD action with $N_f=2$ 
clover improved Wilson quarks and Iwasaki 
improved gauge field action
\begin{eqnarray}
  S   &=& S_g + S_q, \\
  S_g &=&
  -{\beta}\sum_{x,\mu\nu}  \left(
   c_0 W_{\mu\nu}^{1\times1}(x)
   + c_1 W_{\mu,\nu}^{1\times2}(x) \right), \\
  S_q &=& \sum_{f=u,d}\sum_{x,y}\bar{\psi}_x^f \Delta_{x,y}\psi_y^f,
  \label{eq:action}
\end{eqnarray}
where $\beta=6/g^2$, $c_1=-0.331$, $c_0=1-8c_1$,
\begin{eqnarray}
 \Delta_{x,y} &=& \delta_{xy} -\kappa\sum_{i=1}^3 \{(1-\gamma_{i})U_{x,i}\delta_{x+\hat{i},y} \nonumber \\&&
    +(1+\gamma_{i})U_{y,i}^{\dagger}\delta_{x,y+\hat{i}}\}
       \nonumber \\ &&
   -{\kappa} \{e^{a\mu_q}(1-\gamma_{4})U_{x,4}\delta_{x+\hat{4},y}   \nonumber\\ &&
    +e^{-a\mu_q}(1+\gamma_{4})U_{y,4}^{\dagger}\delta_{x,y+\hat{4}}\}   \nonumber \\ &&
   +\delta_{xy}{c_{SW}} \frac{\kappa}{\imath}\sum_{\mu<\nu}\sigma_{\mu\nu}P_{\mu\nu},
\label{eq:fermact}
\end{eqnarray}
where $P_{\mu\nu}$  is the clover definition 
of the lattice field strength tensor,
$c_{SW}$ is the Sheikholeslami–Wohlert coefficient.
All parameters of the action, including $c_{SW}$ 
value were borrowed from the WHOT-QCD collaboration
paper~\cite{Ejiri:2009hq}.

In the following, we abbreviate the normalized canonical partition function, $Z_C(n, T, V)/Z_C(0, T, V)$, as $Z_n$ and study the quark number density
\beq
\label{definition_1}
\rho_q={N_q\over V}={1\over V}{\partial \ln Z_{GC}(\theta,V,T)\over \partial \theta} \;, 
\eeq
where $N_q$ is the net quark number in the lattice volume. At imaginary $\mu_q = \imath \mu_{qI}$, the ensemble average of the quark number density (which is also imaginary, $\rho_q = i \rho_I$) can be well approximated by an odd power polynomial at high temperatures above the RW transition temperature $T_{RW}$, and by the Fourier series at lower temperatures~\cite{Takaishi2009, DELIA2009, Takaishi2015, DELIA2017, Gunther2017, Bornyakov:2016wld}.

In Ref.~\cite{Bornyakov:2016wld} we developed a method to compute $Z_n$ numerically. 
We fitted the lattice data for the quark number density $\rho_I$ to some function thus determining $Z_{GC}(\theta,T,V)$ up to a factor and found
the Fourier transform 
\beq \label{Fourier_1}
Z_C(n,T,V)=N_c \int_{-\pi/N_c}^{\pi/N_c} \frac{d\theta_I}{2\pi}
e^{\imath n N_c \theta_I} Z_{GC}(\theta=\imath\theta_I,T,V)\, 
\eeq 
numerically. It was observed
\cite{Takaishi2015,Bornyakov:2016wld} that at  $T>T_{RW}$, 
i.e. above the RW transition, 
one can fit the lattice data for the quark number density $\rho_I$, 
which is a periodic function with period $2\pi/3$, by the function
\begin{equation}\label{rho_vs_mu}
\hat{\rho}_I = a_1\theta_I - a_3\theta_I^3 + a_5\theta_I^5, \,\,\,\,   -\pi/3  < \theta_I < \pi/3\,, 
\end{equation}
where  $\hat{\rho}_I = \rho_{I}/T^3. $
In addition to the dimensionless variable $\hat \rho_q = \rho_q/T^3$
characterizing the ensemble average of the 
quark number density, 
we will use the variable $\ds \varrho_n = {n\over VT^3}$ 
where $n$ is the net baryon number of a particular
physical state. This variable is helpful to remember that the canonical partition functions $Z_{C}(n,T,V)$ are associated with a definite baryon number density $\varrho_n$. 
  
Using eqs.~(\ref{definition_1}-\ref{rho_vs_mu})
one can numerically compute $Z_n$ at $T>T_{RW}$ for $n$ 
up to some $n_{c}$~\cite{Bornyakov:2016wld}
determined by the condition that 
the values of $Z_n$ at $n > n_{c}$
computed via eq.~(\ref{Fourier_1}) 
have alternating sign which is unphysical
and signals that the fit-function~(\ref{rho_vs_mu}) cannot adequately describe high-density 
($ (\varrho_n > \varrho_{n_c}=n_{c}/(VT^3) $)
contributions to $Z_{GC}(\theta,T,V)$.
It should be noted that $n_{c}$ 
depends on the volume while the corresponding 
(dimensionless) density $\varrho_{n_c}$
is only weakly dependent on the volume and 
is approximately equal to $1.6$ for $T/T_c=1.35$

The reason for such unphysical behavior of $Z_n$ 
(note that all $Z_n$ are to be positive) is that
eq.~(\ref{rho_vs_mu}) imposes unphysical condition 
of a phase transition onto our finite-volume system.
Indeed,  defined by eq.~(\ref{rho_vs_mu}) function is discontinuous at $\theta_I = \pi/3 + 2\pi\,k/3,\,\,k= 0, \pm 1, \pm 2, ...,$.

Note that with $Z_n, n<n_{c}$ obtained in \cite{Bornyakov:2016wld} we are able to reproduce 
the number density $\hat{\rho}_I$ dependence on $\theta_I$  for all $\theta_I$ apart from the vicinity of $\theta_I=\pi/3$.

Thus we have to modify the right-hand side 
of eq.~(\ref{rho_vs_mu}) in the vicinity 
of the RW transition. 
This modification should be volume dependent and 
show the discontinuous behavior eq.~(\ref{rho_vs_mu})
only in the infinite volume limit. To compute 
the quark number density $\rho_{I}$ in the close
vicinity of the transition point $\theta_I=\pi/3$ 
is a formidable task. For this reason, 
we follow another strategy 
and use an approximate analytical 
solution for $Z_n$ 
originally suggested in \cite{RobergeWeiss}.
In our study we considerably improve 
the approximation 
suggested in \cite{RobergeWeiss} and demonstrate 
that this improved approximation 
gives rise to the expected behavior of the density
near $\theta_I=\pi/3$ in the finite volume
and produces the discontinuity at this value 
of $\theta_I$ in the infinite volume limit. 

Roberge and Weiss \cite{RobergeWeiss} obtained 
an approximate expression for $Z_n$ in the case 
when $a_5=0$. They computed the quantities
\beq\label{Fourier_2}
Z_{nA} = \frac{\int_{-\imath \pi/3}^{\imath \pi/3}d\theta e^{-\nu F_n(\theta) }}{ \int_{-\imath \pi/3}^{\imath \pi/3} d\theta  e^{-\nu F_0(\theta) }}
\eeq
where\footnote{If simulations on the $N_s^3\times N_t$ lattice are used to determine the quark number density then $\ds \nu={N_s^3\over N_t^3}$.} $\nu=VT^3$ and  
\beq
F_n(\theta)=  -\varrho_n \theta + \frac{1}{2} a_1 \theta^2 + \frac{1}{4} a_3 \theta^4 
\eeq
using the stationary phase method. It should be noticed  that the contour of integration in (\ref{Fourier_2})
is the segment of imaginary axis, however, to reach the saddle point 
$\theta_0$ which lies on the real axis one should deform this contour to
$\ds -\imath {\pi\over 3} \to \theta_0-\imath {\pi\over 3} \to \theta_0 + \imath {\pi\over 3} \to \imath {\pi\over 3} $.
It is reasonable to apply the method in the $V \to \infty$ limit. The stationarity condition has the form
\beq
\label{eq:polynomial2}
 \varrho_n -  ( a_1 \theta + a_3 \theta^3 ) =0
\eeq
with the solution  $ \theta = \theta_0(\varrho_n)$.
Asymptotic expansion of the integral 
in the numerator of eq.~(\ref{Fourier_2})
has the form ($\nu \to\infty$)
\beq\label{eq:pereval_result}
Z_C\left(\nu\varrho_n,T,V\right)=\sqrt{2\pi\over \nu F_n''(\theta_0)} \exp \left[ -\nu F_n(\theta_0) \right] 
\left(1+\sum_{k=1}^{\infty} {\beta_k\over \nu^k}\right) \;,  
\eeq

where the algorithm of determination of 
the coefficients $\beta_k$ 
is described e.g. in \cite{Lavrentev:1987}.

The eq.~(\ref{eq:polynomial2}) can be solved by radicals: 
\begin{equation}\label{eq__mu_on_rho_poly}
\theta_0= \sqrt{\frac{a_1}{3a_3}} \Big[ ( \sqrt{x^2+1} + x )^{1/3} - ( \sqrt{x^2+1} - x )^{1/3} \Big],
\end{equation}
where $\ds x = \varrho_n \frac{\sqrt{27a_3}}{2a_1^{3/2}}$.
Then the approximate expression obtained in \cite{RobergeWeiss} is
\beq\label{result1}
Z_{nA} = \frac{e^{-\nu F_n(\theta_0) } }{ e^{- \nu F_0(\theta_0) }}
\qquad \mbox{where} \qquad
F_n(\theta_0) = - \frac{1}{4} ( a_1 \theta_0^2 -3 \varrho_n \theta_0 )\;.
\eeq
This approximation consists in omitting the pre-exponential factor as well as the series in braces in Eq.~(\ref{eq:pereval_result})

\begin{figure}[ht]
\center 
\includegraphics[width=0.65\linewidth]{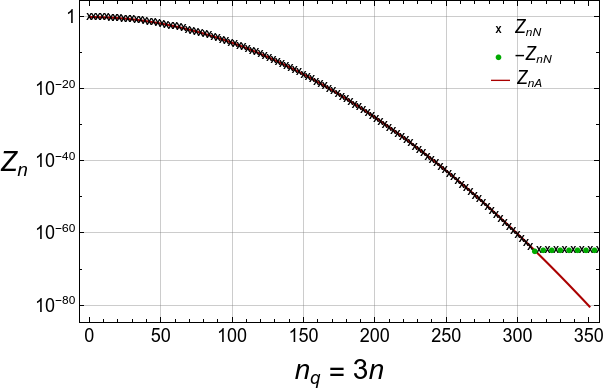}
\caption{$Z_{nN}$ 
(blue crosses) are compared with $Z_{nA}$ (red line) for $N_s=16$. For $n_q>309$ 
positive values of  $Z_{nN}$ are shown by crosses, whereas for negative $Z_{nN}$ we depict  $-Z_{nN}$ using green circles.}

\label{fig:t1_35_Zn_align}
\end{figure}

Thus the normalized canonical partition functions $Z_{n}$ can be computed using the asymptotic approximation Eq.~(\ref{result1}) or using Eq.~(\ref{Fourier_1}) numerically. The canonical partition functions $Z_{n}$, determined by the former procedure, are denoted $Z_{nA}$ by the latter --- $Z_{nN}$. The values of $Z_{nN}$ were computed in~\cite{Bornyakov:2016wld} using the quark number density $\hat{\rho}$ obtained in numerical simulations at $T/T_c=1.35$ on lattices $4 \times 16^3$ and fitted with eq.~(8).
These values and the values of $Z_{nA}$ computed for the respective constants $a_1,~a_3$ are shown in Fig.~\ref{fig:t1_35_Zn_align}. 
One can see that they agree for $n < n_{c}$.  At $n > n_{c}$ $Z_{nN}$ starts to oscillate as described above.

To compare $Z_{nN}$ and $Z_{nA}$ more explicitly, we compute the relative deviation,  
$$
R=(Z_{nN} - Z_{nA})/Z_{nN},
$$ 
for three volumes at $\varrho_n \leq 1.6$.
We find that at $\varrho_n < 1.5 $, the relative deviation is rather small ($R<0.04$), but it increases rather fast, reaching 0.2 
at $\varrho_n = \varrho_{n_c}$. We shall note that the numerical values for 
$a_1=4.671(14)$ and $a_3=0.992(25)$  were obtained on 
$N_s=16$ lattices, and we used the same values for all volumes assuming a very weak volume dependence of $a_i$. Our numerical results on volume dependence of $a_i$~\cite{Bornyakov:2016wld} support this.
Thus, by volume dependence, we will understand the dependence of various observables on $N_s$
through eq.~(\ref{eq:pereval_result}). 
 We checked that variation of $a_i$ within error bars given above did not alter any of our conclusions, although respective variation of $Z_{nA}$ for large $n$ was quite substantial.

In fact, the expression (\ref{result1}) is not the whole story. One can take into account the fluctuations around $\theta_0$ to obtain
\beq
Z_{nA} = \frac{e^{-\nu F_n(\theta_0) } }{ e^{- \nu F_0(\theta_0) }} \frac{\sqrt{F_0^{\prime\prime}(\theta_0)}}{\sqrt{F_n^{\prime\prime}(\theta_0)}}
\label{result3}
\eeq
where 
\beq
F_n^{\prime\prime}(\theta_0) = a_1 + 3 a_3\theta_0^2\;. 
\eeq
Eq.~(\ref{result3}) corresponds to the next-to-leading approximation in the asymptotic expansion (\ref{eq:pereval_result}) consisting in taking into account the pre-exponential factor. In principle, higher-order approximations can also be taken into account. However, we leave this for future 
studies. We find that our result eq.~(\ref{result3}) 
is much closer to $Z_{nN}$ than the result  eq.~(\ref{result1}) obtained in \cite{RobergeWeiss}. 
In this case, $R$ goes to zero in the infinite volume limit for all density values $\varrho_{max}$ under consideration. 
\begin{figure}[htb]
\centering
\hspace*{-0.5cm}
\includegraphics[width=0.56\textwidth,angle=0]{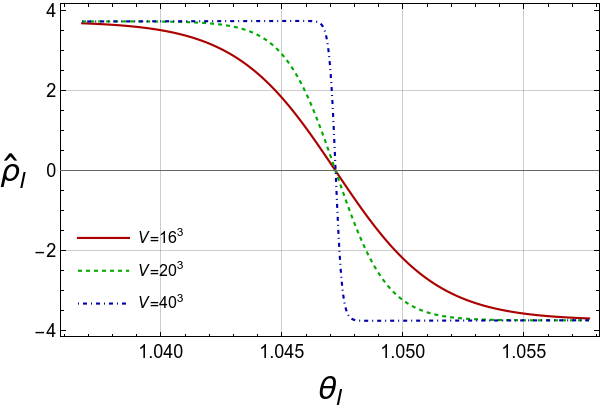}
\vspace{-0.0cm}
\caption{The quark number density computed using $Z_{nA}$ 
for imaginary chemical potential for three volumes and $\varrho_{max}=8.0$ over a restricted range near $\theta_I=\pi/3$.}
\label{dens_im}
\end{figure}
\begin{figure}[htb]
\centering
\hspace*{0cm}
\includegraphics[width=0.48\textwidth,angle=0]{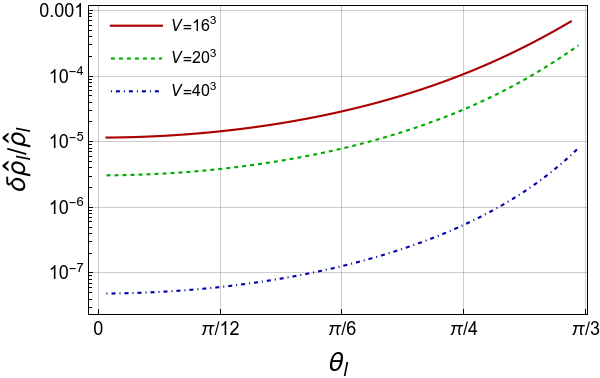}
\includegraphics[width=0.48\textwidth,angle=0]{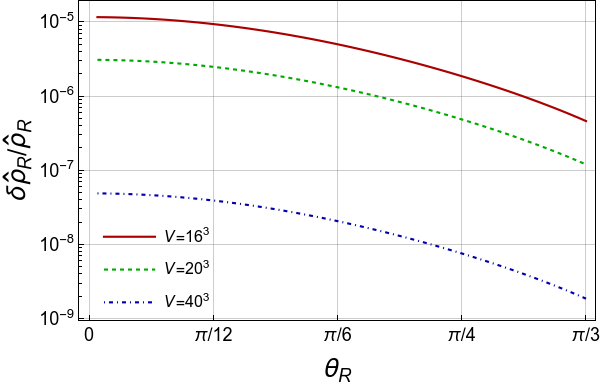}
\vspace{-0.2cm}
\caption{Relative deviation of the quark number density computed using $Z_{nA}$ 
for the imaginary chemical potential (left panel) and the real chemical potential (right panel).
}
\label{reldev_dens_im}
\end{figure}

Another way to check the quality of the obtained
approximation for $Z_n$ is to compute 
the quark number density using the following relations:
(see eqs. (5)-(6) in \cite{Bornyakov:2016wld}):
\begin{eqnarray}
\label{rho_I}
\hat{\rho}_I & = & {1\over \nu} \frac{2 \sum_{n>0} n Z_{nA} \sin (n\theta_I)}{1 + 2 \sum_{n>0}  Z_{nA} \cos (n\theta_I)}\,,\\
\label{rho_R}
\hat{\rho}_R & = & {1\over \nu} \frac{2 \sum_{n>0} n Z_{nA} \sinh (n\theta_R)}{1 + 2 \sum_{n>0}  Z_{nA} \cosh (n\theta_R)}\,
\end{eqnarray}
and compare it with the 
input number density eq.~(\ref{rho_vs_mu})
or with its analytical continuation to real values 
$\theta_R$.
In Fig.~\ref{dens_im} 
we show the quark number density $\hat{\rho}_{I}$ 
computed via  $Z_{nA}$
in the vicinity of the critical value $\theta_I=\pi/3$. 
The volume dependence of the density becomes visible: with increasing volume, the number density approaches a step function, i.e. the first order phase transition appears in the infinite volume limit. 

The relative deviations between results obtained for the quark number density  via Eqs.~(\ref{rho_I}), (\ref{rho_R}) and Eq.~(\ref{rho_vs_mu})  are shown in Fig.~\ref{reldev_dens_im} 
for three spatial volumes. 
In the left panel, it is shown for imaginary $\mu_q$.  
The deviation is small and shows a very fast decrease with increasing $V$.
Note that the relative deviation increases near $\theta_I=\pi/3$ since for this range Eq.~(\ref{rho_vs_mu}) gives a correct number density in the infinite volume limit only. 
In the right panel, we show the relative deviation for 
the real chemical potentials. Again we see that this
relative deviation is small and decreases with volume increase.
Moreover, one can see decreasing with increasing chemical potential.

For $a_5\neq 0$, an explicit solution of eq.~(\ref{rho_vs_mu}) is not available. However, we find an approximate solution considering $a_5$ as a small parameter. Details of the computation are presented in Appendix.
Since $a_5=0$ within error bars at $T=1.35 T_c$ considered here, we estimated the corrections due to higher-order terms 
at lower temperature $T=1.2 T_c$.
We found a good qualitative agreement between $Z_{nN}$ and $Z_{nA}$ for  $n<n_{c}=291$: the relative deviation $|R|<0.04$ 
for $n < 200$  and it runs up to $0.4$ as $n$ increases up to 290. 

However, we should comment on the limitations of the asymptotic approximation eq.~(\ref{eq:pereval_result}). 
If $n>N_{tot}=2N_f N_s^3$,  where $N_{tot}$ is the total number of quark modes on the lattice under consideration,\footnote{Note that there also exist  $N_{tot}$ antiquark modes.}then lattice regularization yields $Z_n=0$, whereas $Z_{nA} \neq 0\ \ \forall n\;$.
Therefore, the range of validity of the approximation~(\ref{Fourier_2}) is 
\begin{equation}\label{eq:range_validity}
 \varrho_n <\!\!\!< \varrho_{tot} 
 = {N_{tot} \over VT^3} 
 = 2 N_f N_t^3\;.
\end{equation}
The values of $\varrho_n$ used in our study are well below the indicated limit.

\section{Evaluation of Lee-Yang Zeros}
\label{sec:others}

Lee-Yang zeros are zeros of the grand canonical partition function $Z_{GC}(\theta, T, V)$, considered a polynomial of the baryon fugacity $\xi_B$. 
In the finite volume $V$, $Z_{GC}(\theta, T, V)$ can be presented as 
\begin{equation}
Z_{GC}(\theta,T,V) =
\sum_{n=-N_{tot}}^{N_{tot}} Z_{C}(n,T,V)\xi_B^n = 
e^{-N_{tot}N_c \theta} \sum_{n=0}^{2 N_{tot}} Z_{C}(n - N_{tot},T,V)\xi_B^n\;,
\label{Eq:FugacityExpansion2}
\end{equation}
Computation of the high-degree polynomial zeros, Eq.~(\ref{Eq:FugacityExpansion2}), is challenging. 
In this work, we employ a very efficient package,  MPSolve v.3.1.8 (Multiprecision Polynomial Solver)~\cite{MPsolve1, MPsolve2}, which provides calculation of polynomial roots with arbitrary precision.

In eq.~(\ref{Eq:FugacityExpansion2}) we rewrote $Z_{GC}(\theta, T, V)$ as a polynomial of degree $2N_{tot}$  which roots represent the values of the LYZ in the complex $\xi_B$-plane. 
Coefficients $Z_{C}(n,T,V) $  are computed using Eq.~(\ref{result3}) with the GNU MPFR Library for multiple-precision
floating-point computations. 
We performed calculations using the MPFR library with accuracy from  1000 to 6000 digits. We found that 1000 digits are enough for the MPSolve package to calculate LYZ for a polynomial with a degree less than 20000. We also checked that to compute LYZ for polynomials of degrees larger than 60000, it is necessary to use $Z_{nA}$ calculated with an accuracy of more than 6000 digits.

To make a conclusion about the infinite-volume limit, we have to study the dependence of the LYZ on $N_s$ and on the number of terms used in the sum eq.~(\ref{Eq:FugacityExpansion2}), which we denote as  $N_{max}$ in what follows.
Analytical expression (\ref{result3}) for the coefficients of the polynomial in Eq.~(\ref{Eq:FugacityExpansion2})
together with the use of the MPSolve package, allow us to work with an arbitrary degree of the polynomial limited only by the computational cost of  LYZ calculation and to improve our previous results~\cite{Wakayama:2018wkc} substantially for $T>T_{RW}$. 
Our main goal is to understand the properties of LYZ in the vicinity of the RW  transition at $\theta_I=\pi/3$.

\begin{figure}[h!]
\center \includegraphics[width=.7\linewidth]{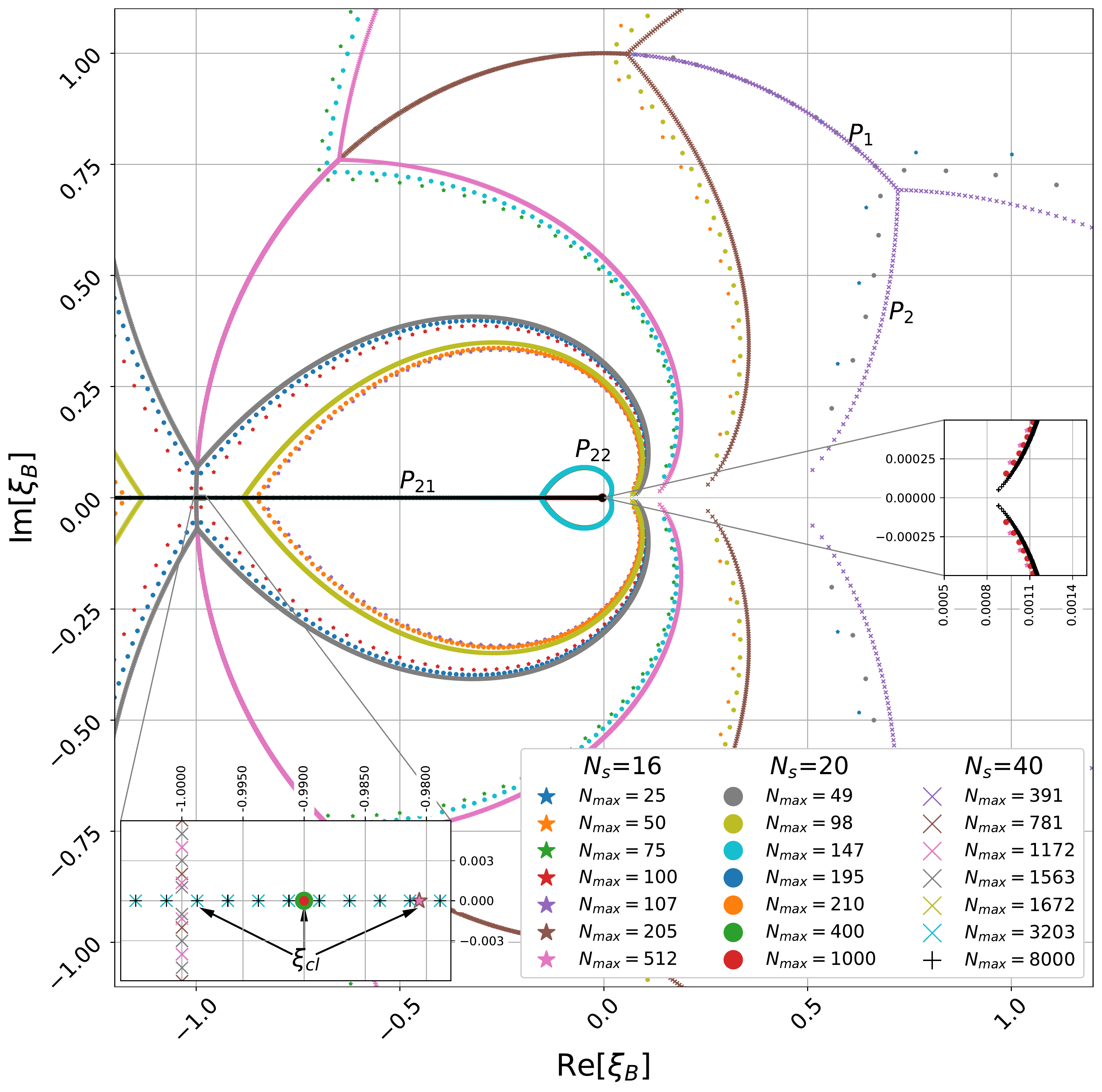}
\caption{LYZ distribution in the fugacity plane. See explanation in the text.
}
\label{fig:fullLYZdistrib}
\end{figure}

In Fig.~\ref{fig:fullLYZdistrib} the results for LYZ are presented in the fugacity plane for three volumes with $N_s=16, 20, 40$ and six values of $N_{max}$ for each volume.
These values of $N_{max}$ change with volume and correspond to six volume independent values of the maximal quark number density $\varrho_{max} = N_{max}/(VT^3)$. 
We start with  $\varrho_{max} \approx 0.39$, and increase it up to $\varrho_{max} = 8.0$.
Note that below we discuss only LYZ located on and within the unit circle of the fugacity plane since the rest of LYZ can be restored using the property $Z_{GC}(\xi_B) = Z_{GC}(1/\xi_B)$.

It was found in Ref.~\cite{Wakayama:2018wkc} that the dependence on $\varrho_{max}$ is much more pronounced than the dependence on the volume. For small $\varrho_{max}$ LYZ consist of two parts (see Fig.~\ref{fig:fullLYZdistrib}): one part, $P_1$, comprises the LYZ distributed along the unit circle  $|\xi_B| = 1$ and the other, $P_2$, represents a curve starting on the circle at the end point of $P_1$ and extending towards the real positive axis. With increasing volume, the endpoint of $P_2$  tends to approach the real axis. 

It was demonstrated in \cite{Wakayama:2018wkc} that $P_1$ shrinks to $\xi_B = -1$ with increasing $\varrho_{max}$. 
Respectively, one end of $P_2$ moves along the unit circle 
towards $\xi_B = -1$ while the other end slowly moves in the direction of $\xi_B = 0$.  
We now confirm these observations, as can be seen from the comparison of LYZ for $\varrho_{max} = 0.39, 0.78, 1.17, 1.56$ in Fig.~\ref{fig:fullLYZdistrib}. 
Moreover, while only the range $\varrho_{max} < 1.6$ was studied in Ref.~\cite{Wakayama:2018wkc} due to the problem with computation of $Z_n$, here we present results for higher values $\varrho_{max} = 1.67, 3.20, 8.0$. 
As a result, we discover the correct properties of the LYZ as described below.  

In Fig.~\ref{fig:fullLYZdistrib}  one can see that $P_1$ disappears completely at $\varrho_{max} > \varrho_{n_c} \approx 1.6$. It is worth noting that this is approximately the same value of the quark density which is reachable in computations of $Z_{nN}$. 
It is natural to assume that the leading contributions to the grand canonical partition function come from all densities over the range $\varrho_n \leq \varrho_{n_c}$ and thus neglecting even part of them gives rise to unphysical results for the LYZ (the $P_1$ part 
on the unit circle).

It should be noticed that the first real negative LYZ also appears when $\varrho_{max}$ becomes as high as $\varrho_{n_c}$  for all volumes under consideration.

The $P_2$ line splits into two parts for $\varrho_{max} > \varrho_{n_c}$ $P_{21}$ and $P_{22}$. $P_{21}$ consists of LYZ on the segment $ -1 < \xi_B < - \xi_{B0}$ with  $\xi_{B0}>0$ moving toward zero exponentially fast with increasing  $\varrho_{max}$. 
The curly part $P_{22}$ connects $-\xi_{B0}$ with the point slightly above the positive real axis (see right insertion in Fig.~\ref{fig:fullLYZdistrib}).  
When we first increase $\varrho_{max}$ to its maximal value for a given volume (in practical terms, this implies 
the extrapolation to the infinite value) and then take the infinite volume limit, we expect that the right end of  $P_{22}$ goes to $\xi_{B} =0$ and complex roots disappear completely.
In fact, we found out that with increasing $\varrho_{max}$ the region occupied by the complex roots shrinks exponentially fast, but their share decreases rather slowly and remains finite. We believe that this difference between our expectations and real observation is due to artefacts of our approximation $Z_{nA}$.

Thus we see a very strong dependence of LYZ on $\varrho_{max}$. At the same time, as can be seen in Fig.~\ref{fig:fullLYZdistrib} the finite volume effects are rather mild. 
For the maximal value of  $\varrho_{max}$ shown in the figure, one can see these effects in the enlarged insertion only. 

Next, we describe the properties of the LYZ on the real-negative axis. We found that for a fixed volume, they start to appear when $N_{max}$ exceeds some value roughly equal to $\ds \varrho_{n_c} \cdot VT^3$. 
The LYZ most close to $\xi_B=-1$ appear first and, with increasing $N_{max}$, additional LYZ, more and more remoted from -1, also appear. This property of LYZ one can see in Fig.~\ref{fig:fullLYZdistrib} ($P_{21}$ increases its length). 
In the left-bottom insertion in Fig.~\ref{fig:fullLYZdistrib} the interval close to  $\xi_B=-1$ is presented. It is seen that for fixed $N_s$  the position of the LYZ located on the real axes is not depending on $N_{max}$. In this insertion we also introduce the notation $\xi_{cl}$  as the location of the negative real LYZ closest to $\xi_B=-1$. 

We approximate the normalized density of LYZ on the real axis, defined as 
\beq
g(\theta_R)= {1\over N_s^3} \frac{dN_{LYZ}(\theta_R)} {d\theta_R} \,,
\eeq
where $N_{LYZ}(\theta_R)$ is the number of LYZ in the interval between 0 and  $\theta_R$, with
\beq
 g(\theta_R)= \frac{1}{N_s^3\; \Delta \theta_R}\,,
 \eeq
where $\Delta \theta_R$ is the distance between adjacent LYZ  on the real negative semi-axis.
 
\begin{figure}[h!]
{\vspace{-0cm}
\hspace{-2cm}
\includegraphics[width=0.5\linewidth]{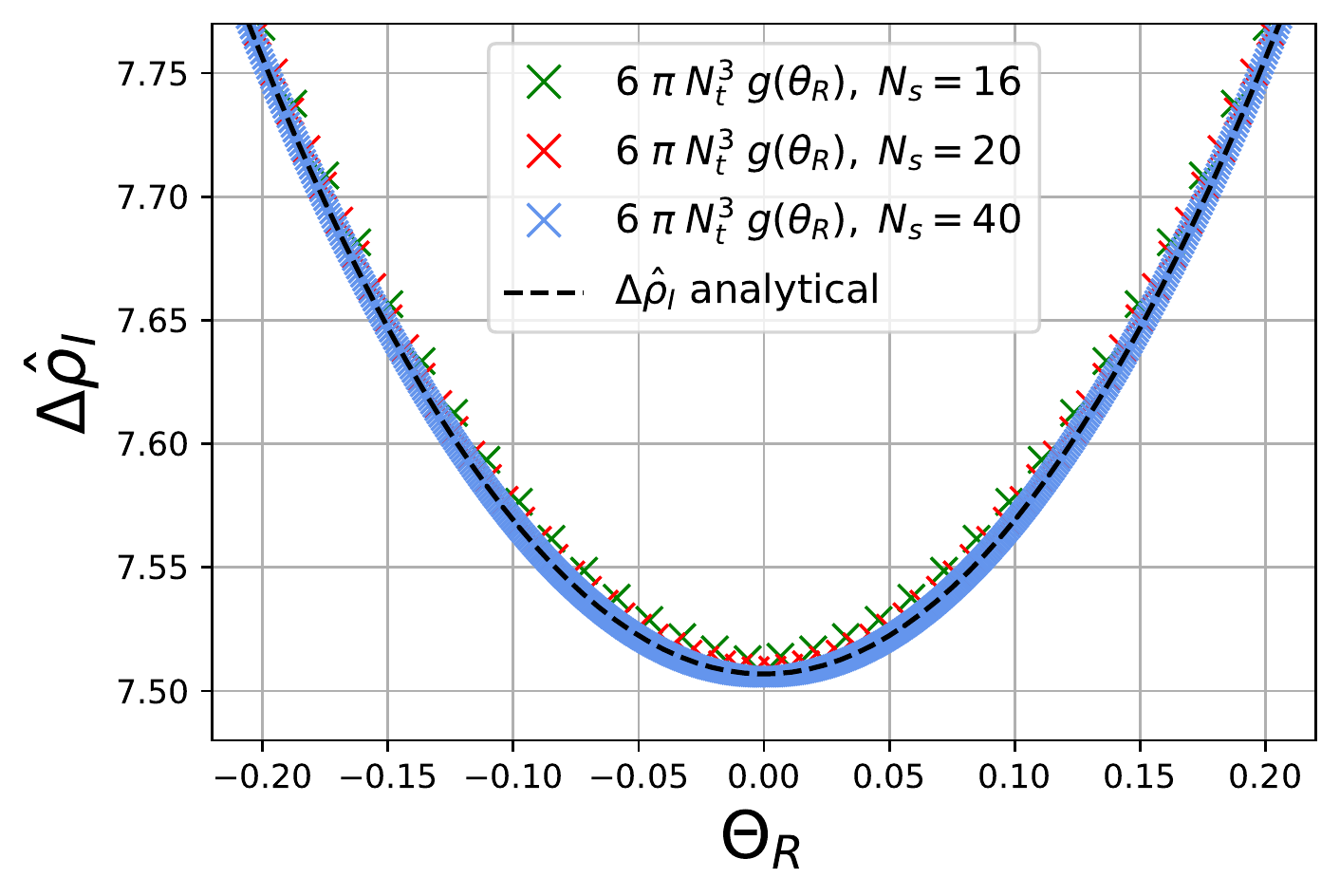}}
\hspace{-0cm}
\includegraphics[width=0.5\linewidth]{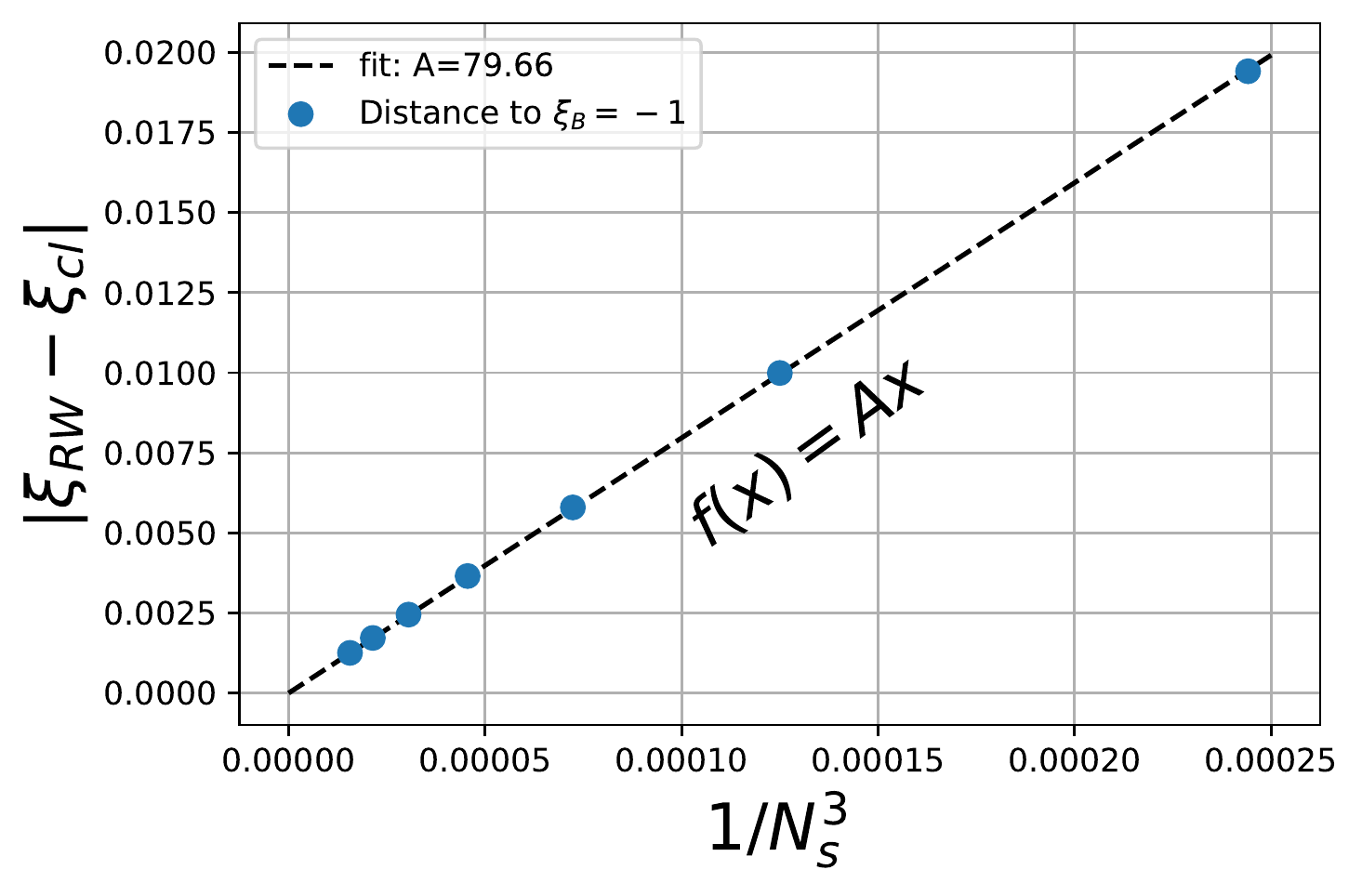}
\caption{ Left: lhs and rhs (for $N_s=16, 20, 40$) of eq.~(\ref{discontinuity}) vs. $\theta_R$ 
at $\ds \theta_I=\pi/3$. Right: $N_s$ dependence of the distance between the point $\xi_B=-1$ and the nearest LYZ.}
\label{fig:fullLYZdistrib-2}
\end{figure}

The formula relating the LYZ density to the discontinuity in the average particle-number density was obtained in~\cite{Lee:1952ig}. Analogously, for the quark number density $\hat\rho_I$, it is straightforward to show that its discontinuity $\Delta\hat \rho_I$ is related to the density of LYZ on the real axis as
\begin{equation}
    \Delta\hat \rho_I= 2\pi N_c N_t^3 g(\theta_R)\,.
\label{discontinuity}
\end{equation}  
It should be noted that this formula is exact only in the infinite-volume limit.
Taking Eq.~(\ref{rho_vs_mu}) continued to the complex plane one obtains for the discontinuity $\Delta\hat \rho_I$ at $\ds \theta_I={\pi\over 3}$ the following dependence on $\theta_R$:
\beq
\Delta\hat \rho_I = 2 \Big{(}a_1 \frac{\pi}{3} + a_3 \frac{\pi^3}{27}   - a_3 \pi  \theta_R^2 \Big)
\eeq
We show in the left panel of Fig.~\ref{fig:fullLYZdistrib-2} both parts of eq.~(\ref{discontinuity}) as functions of $\theta_R$, results for the right hand side are presented for $N_s=16,20,40$. Dependence on $N_S$ is visible only in the vicinity of $\theta_R=0$.
One can see that  Eq.~(\ref{discontinuity})  is nicely satisfied (the relative deviation for $N_s=40$ is below 0.003) indicating once more that our approximation $Z_{nA}$ for the canonical partition functions works extremely well.  Note that we show in this figure also the LYZ for positive $\theta_R$ corresponding to $|\xi_B| > 1$. The symmetry seen in the figure  demonstrates the property $Z_{GC}(\xi_B) = Z_{GC}(1/\xi_B)$ mentioned above.

Thus we conclude that in the infinite-volume limit, the density of LYZ does not vanish over the entire negative real semiaxis in the fugacity plane, including the point $\xi_{RW} = -1$. According to the criterion suggested in~\cite{Lee:1952ig}, the phase transition occurs at $Im(\mu_B/T)=\pi$.  
Yet another piece of evidence for our conclusion comes from 
$N_s$ dependence of the distance $|\xi_{RW}-\xi_{cl}|$, where $\xi_{cl}$ is the LYZ closest to $\xi_{RW}$. 
This dependence is shown in the right panel of Fig.~\ref{fig:fullLYZdistrib-2}, where results for $N_s=16,20,24,28,32,36,40$ are depicted.
The constant $A$ of the fitting function provides an estimate of the LYZ density:
$ A = \frac{1}{2} g(0)$.
Analogous results can be obtained for other values of $\xi$ on the negative real semi-axis.

\section{Conclusions}
\label{sec:conclusions}
We have studied $N_f=2$ lattice QCD in the deconfinement phase above $T_{RW}$ at $T/T_c=1.35$. 
Our goal was to compute the distribution pattern of LYZ  in the complex fugacity $\xi_B$ plane and to demonstrate the existence of the LYZ corresponding to the RW transition at imaginary quark chemical potential $\mu_{qI}/T = \pi/3$.

In our earlier work \cite{Wakayama:2018wkc} we had performed computations of LYZ at $T>T_{RW}$ using a limited number of  
canonical partition functions $Z_{nN}$ computed numerically via the Fourier transform. 
We then found that $Z_{nN}$ could be computed only for a restricted range of $n$.
To overcome this restriction, we have used the approximate analytical expression $Z_{nA}$ in this work.
We demonstrated that $Z_{nA}$ agree very well with values of $Z_{nN}$ where the latter are available and reproduce the input expression for the quark number density. In both cases, the agreement improves with increasing volume. 
Using $Z_{nA}$, we have obtained the behavior for the quark number density in the vicinity of the RW  phase transition at
$\theta_I=\pi/3$ which is consistent with first principles. 
We shown that, in the infinite volume limit, there 
appears a discontinuity in the dependence of 
$\hat{\rho_I} $ on $\theta_I$ indicating the first-order transition behavior.

After we ensured that $Z_{nA}$ works very well, we computed the LYZ. Apart from increasing the available maximal density $\varrho_{max}$ which is now restricted only by the computation 
resources we also used a more effective procedure to compute LYZ, which employs the MPSolve library with arbitrary precision. 
Our computations have been performed with precision up to 6000 significant digits. 

We confirmed our previous results about the strong dependence of LYZ on $\varrho_{max}$ and comparatively weak volume dependence, which were obtained in~\cite{Wakayama:2018wkc} with restricted $\varrho_{max}$.

Then, increasing $\varrho_{max}$, we have discovered that the LYZ appear on the $\xi$ negative real half-axis. In the infinite volume limit, LYZ tend to fill the full $Re(\xi) \leq 0$ range and have a non-vanishing density corresponding to the RW 
phase transition.
We demonstrated that in the limit of high $\varrho_{max}$ LYZ exist away from the $\xi$ negative real axis, but only in the tiny vicinity of $\xi=0$ point. 
We believe that these LYZ  are artefacts of our approximation  $Z_{nA}$.

We derived the relation between the quark number density discontinuity $\Delta \rho_I$ and the LYZ density eq.~(\ref{discontinuity}) and observed that our numerical results for the LYZ density nicely satisfy this relation (see Fig.~\ref{fig:fullLYZdistrib-2}, left panel). This agreement demonstrates once more that approximation  $Z_{nA}$ works extremely well.

Our result demonstrates the necessity of using sufficiently high values of $\varrho_{max}$  to obtain correct results for LYZ. 
We believe that our results, obtained at a high temperature where phase transitions are absent at real chemical 
potential, will be useful for the computation of LYZ at a low temperature where phase transition or crossover at real $\mu_B$ is present. This work is now underway.

\acknowledgments{This work was supported by the Russian Foundation for Basic Research via grant 18-02-40130 mega
and partially carried out within the state assignment of the Ministry of Science and Higher Education of Russia (Project No. 0657-2020-0015).
Computer simulations were performed on the FEFU GPU cluster Vostok-1, the Central Linux Cluster of the 
NRC ”Kurchatov Institute” - IHEP, the Linux Cluster of the KCTEP NRC ”Kurchatov Institute”. In addition, we used computer resources 
of the federal collective usage center Complex for Simulation and Data Processing for Mega-science Facilities at NRC "Kurchatov Institute",
http://ckp.nrcki.ru/. Participation of  Denis Boyda at the earlier stages of this work is gratefully acknowledged.}

\appendix
\section{Corrections for $a_5\neq 0$.}

The approximate solution 
$\theta$ of the equation 
\beq 
a_5 \theta^5 + a_3 \theta^3 + a_1 \theta - \varrho_n =0
\eeq 
up to the order $a_5^2$ has the form 
\begin{equation}
\theta= \theta_0 + a_5\ \theta_1 + a_5^2\ \theta_2,
\end{equation}
where 
$\theta_0$ is defined in eq.~(\ref{eq__mu_on_rho_poly}),
\begin{eqnarray}
\theta_1 & = & -\frac{\theta_0^5}{a_1 + 3 a_3 \theta_0^2}\;, \\
\theta_2 & = & - \frac{5\theta_1 \theta_0^4 + 3 a_3 \theta_1^2 \theta_0}{a_1 + 3 a_3 \theta_0^2}\;.
\label{a5_pert1}
\end{eqnarray}
Then we follow the procedure described in eqs.~(\ref{Fourier_2}-\ref{result3}) and obtain
\beq
Z_{nA} = \frac{e^{-\nu F_n(\theta) } }{ e^{- \nu F_0(\theta) }} \frac{\sqrt{F_0^{\prime\prime}(\theta)}}{\sqrt{F_n^{\prime\prime}(\theta)}}\,,
\label{Zn_1p20}
\eeq
where
\begin{equation}
F_n(\theta) 
= - \frac{1}{4} ( a_1 \theta_0^2 -3 \varrho_n \theta_0  + a_5 f_1(\theta_0) + a_5^2 f_2(\theta_0) + O(a_5^3))\,,
\label{a5_pert2}
\end{equation}

\beq
F_n^{\prime\prime}(\theta) = 
 a_1+3a_3\theta_0^2 + a_5 f_3(\theta_0) + a_5^2 f_4(\theta_0) + O(a_5^3)
\label{a5_pert3}
\eeq
and
\begin{subequations}
\begin{eqnarray}
f_1( \theta_0 ) & = & \frac{2}{3} \theta_0^6\,,\\
f_2( \theta_0 ) & = & - \frac{2 \theta_0^{10}}{a_1 + 3 a_3 \theta_0^2} \,,\\
f_3( \theta_0 ) & = & \theta_0^4 \left( 3 + \frac{2 a_1}{a_1 + 3 a_3 \theta_0^2} \right) \,,\\
f_4( \theta_0 ) & = & - \theta_0^8 \frac{ 20 a_1^2 + 87  a_1 a_3 \theta_0^2 + 99 a_3^2 \theta_0^4  }{(a_1 + 3 a_3 \theta_0^2)^3}\,.
\end{eqnarray}
\end{subequations}

To compute $Z_{nA}$ for $T/T_c = 1.20$ we use  
$a_1=4.409, a_3=1.032, a_5=-0.165$  \cite{Bornyakov:2016wld} 
in eq.~(\ref{rho_vs_mu}). It should be noted 
that with this set of  constants $a_i$   eq.~(\ref{rho_vs_mu}) 
is not applicable at large real chemical potential 
since the quark number density  becomes negative
at $\theta>3.028$.
To keep the density positive we should assume 
that the coefficient $a_7$ which 
was not determined in  \cite{Bornyakov:2016wld} is positive.

\end{document}